\documentstyle[preprint,aps]{revtex}
\begin{document}
\draft
\def \a1{{\alpha_1}}
\def \a2{{\alpha_2}}
\def \v{{\vec u}}
\preprint{MRI-PHY/P980552}
\title{ Non-Adiabatic Distortion In The Current Distribution\\
Around A Moving Vortex}
\author{D.M. Gaitonde}
\address{Mehta Research Institute,\\ 
          Chatnaag Road, Jhusi,\\
             Allahabad 211019, INDIA. \\ }
\maketitle
\begin{abstract}
We study the phase distribution around a vortex in uniform
motion. We consider both the cases of neutral and
charged superfluids. The motion of the vortex causes
the density of the system to fluctuate.
This in turn produces a compensating current, 
thus ensuring current conservation locally.
We explicitly calculate this current.
\end{abstract}

\newpage

In recent years, there has been a great deal of interest (see Ref. 1
for a recent review) in the static and dynamic properties
of vortices in superfluids and superconductors.
A study of  high temperature and other superconductors
has led to the discovery of several new phenomena
such as flux-lattice melting (2), quantum flux creep (3),
anomalous sign change of the Hall co-efficient (4)
and anomalous a.c. electromagnetic response(5).
This has led to a renewed upsurge in efforts to understand
the dynamics of moving vortices. Issues like the size
and sign of the Magnus force (6) and the vortex inertial mass
(7)have been intensely studied.
Theoretical studies of these phenomena have been carried out, 
both within an effective Ginzburg-Landau theory (8) and 
microscopically (9) within the BCS theory.

However, most studies of vortex dynamics have tended
to assume the vortex motion to be adiabatic i.e.
the instantaneous distribution of the superconducting order
parameter around the vortex is assumed to be identical
with that of a vortex at rest.
In this paper we address the question of non-adiabatic 
distortions in the phase distribution around a moving
vortex. Our calculations are performed using a 
Ginzburg-Landau phase-only functional. 
We solve the equations of motion to second
order in the vortex velocity ($u$)  and thus obtain
corrections to the phase distribution.

The vortex motion induces fluctuations in the instantaneous
density distribution around the vortex.
Current conservation therefore causes a compensating
current to flow, thus conserving the density
locally. We explicitly calculate the induced current,
to second order in $u$ for neutral as well as 
charged superfluids. The constant current contours
are found to be distorted from their circular
shape. However, there is no change in the energy,
to second order in $u$, coming from the non-adiabatic distortions
in the case of a neutral superfluid. 
For a charged superfluid the displacement current,
associated with a time-dependent vector potential,
contributes to a change in the energy.
This contribution is suppressed by a factor of
$u^2/c^2$ and is therefore negligibly small.
These effects, thus have no consequences either
for the Magnus force or for the vortex mass.

The superfluid is described by the action functional
per unit length,
$S^{\prime}=S/L=\int dt\int d{\vec r} [{\rm L_{\theta}}+{\rm L_{em}}]$ 
where 
$${\rm L}_{\theta}=
\frac{\alpha_1}{2} \left( \dot{\theta} -
\frac{2eA_0}{\hbar} \right)^2 - \frac{\alpha_2}{2}
\left({\nabla}\theta - \frac{2e\vec{A}}{\hbar
c}\right)^2 +\gamma\dot{\theta} \eqno{(1a)}$$
and 
$${\rm L_{em} }= 
 \frac{(\nabla A_0+{1\over c}{\partial\vec{A}\over \partial t})^2 -
({\nabla} \times \vec{A})^2}{8 \pi}  \eqno{(1b)}$$
Here $\theta$ is the phase of the superconducting order
parameter and $A_0$ and ${\vec A}$ are the scalar and vector
potentials associated with the electromagnetic field.
This functional has been derived microscopically (10)
and the co-efficients $\alpha_1$ and $\alpha_2$ can be related
to appropriate polarizabilities of the underlying fermionic system.
The size and sign of the co-efficient $\gamma$, which determines the
Magnus force, remains controversial.
For purposes of this paper, we treat the Ginzburg-Landau
coefficients as phenomenological parameters.

We will first consider the case of a neutral superfluid.
Accordingly, we put $e=0$ in Eq. (1) and switch off 
the electromagnetic fields. We then have,
$$L^{ns}=\frac{\alpha_1}{2}  \dot{\theta}^2 
 - \frac{\alpha_2}{2}({\nabla\theta})^2+\gamma{\dot \theta}\eqno{(2)}.$$
 The Euler-Lagrange equation of motion is easily found to be
 $$\alpha_1 {\ddot{\theta}}=\alpha_2\nabla^2\theta\eqno{(3)}.$$
Notice that the term linear in $\dot\theta$
corresponds to a topological phase and makes no contribution
to the equation
 of motion.

For a uniformly moving vortex we have $\theta({\vec r},t)=\theta({
\vec r}-{\vec u}t)$. Substituting this  in Eq. (3)
we get 
$$\alpha_1 \v\cdot\nabla(\v\cdot\nabla\theta)=\alpha_2
\nabla\cdot(\nabla \theta) \eqno{(4)}$$
 We seek a solution to Eq. (4) of the form
$$ \theta({\vec r},t)=\theta^{(0)}({\vec r}-{\vec u}t)+
\theta^{(1)}({\vec r}-{\vec u}t)+
\theta^{(2)}({\vec r}-{\vec u}t)+....\eqno{(5)}$$
where $\theta^{(m)}({\vec r}-{\vec u}t)$ is of $m^{th}$ order
in $\v$. Here $\nabla\theta^{(0)}({\vec {r^\prime}})=
{\hat{\phi}^{\prime}\over r^{\prime}}$ (where we define the co-ordinate
${\vec {r^\prime}}={\vec r}-{\vec u}t$ for convenience)
is the phase distribution of a static vortex  while the remaining
terms are non-adiabatic phase distortions induced by the vortex motion.
The quantization of the vorticity, which stems from the single-valuedness
of the superconducting order parameter, ensures that for $m\neq 0$
$\theta^{(m)}({\vec {r^\prime}})$ is non-singular and
$\nabla \theta^{(m)}$ is purely longitudinal.

Substituting Eq. (5) in Eq. (4) and equating terms of the same
order in $u$, we arrive at the result
$$\nabla^{\prime 2}\theta^{(1)}({\vec {r^\prime}})=0 \eqno{(6a)}$$
and
$$\alpha_1 \v\cdot  \nabla^{\prime }
(\v\cdot\nabla^{\prime }\theta^{(0)}({\vec {r^\prime}}))
=\alpha_2\nabla^{\prime 2}\theta^{(2)}({\vec {r^\prime}})\eqno{(6b)}$$
Since $\nabla \theta^{(1)}$ is constrained to be purely longitudinal,
Eq. (6a) implies that it is zero. We therefore set 
$\theta^{(1)}=0$.

Fourier transforming Eq. (6b) with respect to ${\vec {r^\prime}}$
we solve for $\theta^{(2)}_{\bf q}$ to get
$$\theta^{(2)}_{\bf q}={-2\pi\alpha_1\over \alpha_2}\v\cdot{\vec q}
{\v\cdot{\hat z}\times{\vec q}\over q^4}\eqno{(7a)}$$
and find the corresponding current $\nabla^{\prime}\theta^{(2)}_{\bf q}$
to be
$$\nabla^{\prime}\theta^{(2)}_{\bf q}=
{2\pi\alpha_1\over \alpha_2}{\vec q}(\v\cdot{\vec q})
{\v\cdot{\hat z}\times{\vec q}\over iq^4}\eqno{(7b)}
$$
It is easy to see from Eq. (2) that this current, which is purely
longitudinal, doesn't contribute to the action to second order
in $u$. 
This implies that both the Magnus force and the vortex mass
are unaffected by the existence of this current.
To get a better idea of the induced current distortion,
we Fourier transform $\nabla^{\prime}\theta_{\bf q}$
(the details are outlined in the Appendix)
to get
$$\nabla^{\prime}\theta({\vec {r^\prime}})={{\hat\phi}^{\prime}\over 
r^{\prime}}[1-{\alpha_1 u^2\over 2\alpha_2} \cos 2\phi^{\prime}]\eqno{(8)}$$
for the x-axis chosen to be directed along ${\vec u}$.
It is easily seen from Eq. (8) that the circular symmetry
of a static vortex is destroyed by the extra current.

We now turn our attention to the charged case.
The Euler-Lagrange equations of motion are now given by
$$\alpha_1{\partial\over \partial t}({\dot \theta}-{2e\over \hbar} A_0)=\alpha_2
\nabla\cdot(\nabla\theta-{2e\over \hbar c}{\vec A}) \eqno{(9a)}$$
$${\nabla\cdot(\nabla A_0+{1\over c}{\partial\vec{A}\over \partial t})\over 4\pi}
=-{2e\alpha_1\over \hbar}
(\dot{\theta}-{2e\over\hbar}A_0)\eqno{(9b)}$$
and
$${\nabla\times(\nabla\times {\vec A})\over 4\pi}
+{1\over 4\pi c}{\partial\over\partial t}(\nabla A_0
+{1\over c}{\partial\vec{A}\over \partial t})=\alpha_2{2e\over \hbar c}(\nabla\theta
-{2e\over \hbar c}{\vec A})\eqno(9c).$$

We choose the gauge $\nabla\cdot{\vec A}=0$. Then, on Fourier
transforming Eq. (9b)  we find
$$A_0({\vec q})={\hbar\over 2e}{\v\cdot\nabla^{\prime}\theta_{\vec q}\over
{q^2\lambda_{TF}^2+1}}\eqno(10)$$
where the Thomas-Fermi screening length is given by
$4\pi\alpha_1({2e\over \hbar})^2=\lambda_{TF}^{-2}$.
On Fourier transforming Eq. (9c) and making use of Eq. (10)
we have
$${\vec A}({\vec q})={(-\v\cdot{\vec q}){\vec q}\hbar\v\cdot\nabla
\theta_{\bf q}\over 2ec (q^2\lambda_{TF}^2+1)(q^2+\lambda_L^{-2}-
(\v\cdot{\vec q}/c)^2)} + {(\hbar c/2e) \nabla \theta_{\bf q}\over
q^2\lambda_L^2+1-(\lambda_L\v\cdot{\vec q}/c)^2}\eqno{(11)}$$
where the penetration depth $\lambda_L$ is given by
$4\pi\alpha_2(2e/\hbar c)^2=\lambda_L^{-2}$.
In writing Eq. (11), we have made use of the relations
${\vec A}({\vec r},t)={\vec A}({\vec r}-\v t)$ and
$A_0({\vec r},t)=A_0({\vec r}-{\vec u}t)$.
All Fourier transforms have been performed with respect
to the co-ordinate ${\vec{r^\prime}}={\vec r}-\v t$ as before.

We now substitute our result for $A_0$ (Eq.10) in Eq. (9a)
and make use of our gauge condition to arrive at
$$\alpha_1\v\cdot {\vec q}{\v\cdot\nabla^\prime\theta_{\bf q}}{
q^2\lambda_{TF}^2\over q^2\lambda_{TF}^2+1}=\alpha_2{\vec q}\cdot
\nabla^{\prime}\theta_{\bf q}\eqno{(12)}.$$
We put $\theta=\theta^{(0)}+\theta^{(1)}+\theta^{(2)}+.....$ as before.
We once again find $\theta^{(1)}$ to be zero and get
$$\theta^{(2)}_{\bf q}=-{2\pi\alpha_1\over\alpha_2}{\v\cdot{\vec q}}{\v\cdot
{\hat z}\times {\vec q}\over q^4}{q^2\lambda_{TF}^2\over q^2\lambda_{TF}^2
+1}\eqno{(13)}.$$
Notice that as $e\rightarrow 0$ and $\lambda_{TF}\rightarrow
\infty$, Eq.(13) recovers the earlier result of Eq. (7a)
for a neutral superfluid.

Substituting Eq. (13) in Eq. (11) we find, to second order
in $u$, the vector potential to be
$${\vec A}({\vec q})={\hbar c\over 2e}{2\pi\over iq^2}{{\hat z}\times
{\vec q}\over q^2\lambda_L^2+1}[1+{\lambda_L^2\over c^2}
{(\v\cdot{\vec q})^2\over q^2\lambda_L^2+1}]\eqno{(14)}.$$
As expected, the vector potential is purely transverse as
the longitudinal terms exactly cancel, consistent with our gauge condition.

Making use of Eqs. (13) and (14) we find that
the current density ${\vec j}({\vec {r^{\prime}}})=[\nabla^{\prime}\theta-{2e\over \hbar c}
{\vec A}] $ is given by ${\vec j}={\vec j}_a+{\vec j}_b+{\vec j}_c$
where 
$${\vec j}_a({\vec q})={2\pi\over iq^2}{{\hat z}\times {\vec q}q^2\lambda_L^2
\over q^2\lambda_L^2+1}\eqno{(15a)}$$
is the cuurent density of a vortex at rest,
$${\vec j}_b({\vec q})={2\pi\lambda_L^2\over i q^2c^2}
{{\hat z}\times {\vec q}q^2\lambda_L^2\over (q^2\lambda_L^2+1)^2}
(\v\cdot{\vec q})^2\eqno{(15b)}$$
corresponds to the "displacement current" of the moving
flux and
$${\vec j}_c({\vec q})={2\pi\lambda_L^2\over i q^2c^2}
{\vec q}({\v\cdot{\vec q}}){\v\cdot{\hat z}\times{\vec q}\over
q^2\lambda_{TF}^2+1}\eqno{(15c)}$$
is the induced current necessary for local charge 
conservation.

We now proceed to evaluate ${\vec j}({\vec {r^\prime}})$.
Fourier transforming ${\vec j}_a({\vec q})$
we find (see Appendix for details)
$${\vec j}_a({\vec {r^\prime}})={{\hat \phi}^{\prime}\over \lambda_L}K_1(r^{\prime}/\lambda_L)
\eqno{(16)}$$
which is the usual result for a vortex at rest.
Here $K_m(x)$ are the modified Bessel functions.
In a similar manner, we evaluate ${\vec j}_b({\vec {r^{\prime}}})$ and
${\vec j}_c({\vec {r^{\prime}}})$, details of which have been
provided in the Appendix.

We find 
$${\vec j}_b({\vec {r^\prime}})={u^2\over 4c^2}[((2+\cos 2\phi^\prime)I_1-\cos 2\phi^\prime
I_2){\hat \phi}^\prime+{\sin 2\phi}^\prime(I_1+I_2){\hat r}^\prime]\eqno{(17)}$$
where
$$I_1={1\over 2\lambda_L}[3K_1(r^{\prime}/\lambda_L) +{r^{\prime}\over\lambda_L}
K_1^{\prime}({r^\prime}/\lambda_L)]\eqno{(18a)}$$
and 
$$I_2={r^{\prime}\over 2\lambda_L^2}K_2({r^\prime}/\lambda_L)
\eqno{(18b)}.$$
Similarly, 
$${\vec j}_c({\vec {r^\prime}})={u^2\lambda_L^2\over 4c^2}[(I_4-I_3)\sin 2\phi^{\prime}
{\hat r}^{\prime}-(I_3+I_4)\cos 2\phi^{\prime}{\hat \phi}^{\prime}]\eqno{(19)}.$$
Here 
$$I_3\approx 0 \eqno{(20a)} ,$$
and
$$I_4 \approx {8\over {r^\prime}^3}(1-J_0(r^\prime/\xi ))-{8\over {r^{\prime}}^2
\xi} J_1(r^\prime /\xi)+ {4\over r^{\prime}\xi^2}J_1^{\prime}(r^\prime /\xi)
\eqno{(20b)},$$
and $J_m(x)$ are Bessel functions.
The smallness of $I_3$ and $I_4$ stems from the small
screening length ($\lambda_{TF}$) associated with
$j_c$ (see Eq.(15c)). This length is much smaller than the
coherence length $\xi$, which is the coarse-graining scale
beyond which a description in terms of a 
 phase-only action functional is applicable.
Notice also that both $j_b$ and $j_c$ are strongly suppressed by factors of
$u^2/c^2$ making their observation a difficult
task.

We now consider the drawbacks of our calculation.
As pointed out by Aitchison et. al. (10), at $T=0$, one has to include
additional terms in the phase-only Lagrangian to ensure Galilean invariance.
However, the solution of the corresponding non-linear problem
is beyond the scope of our work.
We have also ignored effects arising from the vortex core, whose 
proper inclusion would require a use of microscopic theory.

Finally, we summarise the main results of this paper.
We have calculated the non-adiabatic phase distortion 
of a moving vortex.Vortex motion induces time-dependent
density fluctuations, which in turn gives rise 
to additional currents necessary to ensure local
charge conservation.
For the neutral superfluid, the extra supercurrent 
is purely longitudinal 
with observable consequences.
In the charged case
there are two distinct contributions.
One coming from the "displacement current"
($j_b$ above) is purely transverse whereas the other
($j_c$ above) is longitudinal.
These currents are however strongly suppressed 
by factors of $u^2/c^2$ as well as efficient Coulomb screening
in the case of $j_c$, making their observation a 
difficult task.

\newpage
{\bf APPENDIX:}

In this appendix, we outline the derivation 
of the results stated above for the current ${\vec j}({\vec {r^\prime}})$.
We will first take up the case of a neutral superfluid.
Then we have
$$\nabla^\prime \theta_{\bf q}={2\pi {\hat z}\times {\vec q}\over iq^2}
+{2\pi\alpha_1\over\alpha_2}{\vec q} ({\vec u}\cdot {\vec q})
{{\vec u}\cdot {\hat z}\times {\vec q}\over i q^4}\eqno{(A1)}$$
where the first term on the R.H.S. is the usual 
current of a vortex at rest and the second term is
the distortion induced by the vortex motion.
Thus, the current in real space is given by
$$\nabla^{\prime}\theta({\vec {r^\prime}})=\int {d{\vec q}\over (2\pi)^2}\exp{(i {\vec q}\cdot
{\vec {r^\prime}})}\nabla^\prime \theta_{\bf q}\eqno{(A2)}.$$
The first term in Eq. (A1) is easily Fourier transformed
to yield the standard result
$$\nabla^{\prime}\theta({\vec {r^\prime}})_a
={{\hat \phi}^\prime\over r^{\prime}}\eqno(A3).$$
We find, on performing the angular integrals 
in Eq. (A2), the second term in the current to be
$$\nabla^{\prime}\theta({\vec {r^\prime}})_b={-\alpha_1 u^2\over
4\alpha_2}\int dq [{\hat r}^{\prime}\sin {2\theta}(J_3(qr^\prime)
-J_1(qr^{\prime}))+{\hat \phi}^{\prime}\cos {2\theta}(J_3(qr^\prime)
+J_1(qr^{\prime}))\eqno{(A4)}$$
where $\theta$ is the angle $\v$ makes with respect to
${\vec {r^\prime}}$.
The integrals in Eq. (A4) can now be performed to yield
$$\nabla^{\prime}\theta ({\vec {r^\prime}})_b=
{-\alpha_1u^2\over 2 \alpha_2}\cos 2\theta {{\hat \phi}^\prime
\over r^\prime}\eqno{(A5)}.
$$ On adding the results of Eqs. (A3) and (A5)
and choosing ${\vec u}$ to lie along the x-axis
we finally arrive at Eq.(8) above.

We now consider the charged case.
In this case, the current ${\vec j}({\vec q})$ is the sum
of three different contributions:$ {\vec j}_a$,
${\vec j}_b$ and ${\vec j}_c$. The corresponding Fourier
transforms ${\vec j}_a({\vec q})$, ${\vec j}_b
({\vec q})$ and ${\vec j}_c({\vec q})$ have been obtained
earlier (Eq. (15)).
We will evaluate each piece seperately.
$${\vec j}_a({\vec {r^\prime}})=\int {d{\vec q}\over( 2\pi)^2}
\exp{(i{\vec q}\cdot
{\vec {r^{\prime}}})}{2\pi\over i q^2}{{\hat z}\times {\vec q}q^2
\lambda_L^2\over q^2\lambda_L^2+1}\eqno{(A6)}$$
is the usual current associated with a static vortex.
On performing the angular integration we get
$${\vec j}_a({\vec {r^\prime}})={\hat \phi}^{\prime}\int dq {q^2 J_1(qr^\prime )\over
q^2+\lambda_L^{-2}}={{\hat \phi}^{\prime}\over \lambda_L}K_1(r^\prime/
\lambda_L)\eqno{(A7)}$$

We now consider ${\vec j}_b({\vec {r^\prime}})$.
$${\vec j}_b({\vec {r^\prime}})=\int {d{\vec q}\over (2\pi)^2}
\exp{(i{\vec q}\cdot
{\vec {r^{\prime}}})}{2\pi\lambda_L^2\over i q^2c^2}
{{\hat z}\times {\vec q}q^2\lambda_L^2\over (q^2\lambda_L^2+1)^2}
(\v\cdot{\vec q})^2\eqno{(A8)}$$
On performing the angular integral we arrive at Eq. (17)
above
where 
$$I_1=\int dq {q^4\lambda_L^4J_1(qr)\over (q^2\lambda_L^2+1)^2}
\eqno {(A9a)}$$
and
$$I_2=\int dq {q^4\lambda_L^4J_3(qr)\over (q^2\lambda_L^2+1)^2}
\eqno{(A9b)}
$$
$I_1$ can be rewritten as
$$I_1={-\lambda_L^3\over 2}{\partial\over \partial\lambda_L}
\int dq {q^2 J_1(qr^\prime)\over q^2\lambda_L^2+1}=
{-\lambda_L^3\over 2}{\partial\over \partial\lambda_L}
[{K_1(r^\prime /\lambda_L)\over \lambda_L^3}]\eqno{(A10)}$$
which reduces to the result of Eq. (18a).
$I_2$ is directly evaluated to yield the result
stated earlier in Eq. (18b).

Finally we consider ${\vec j}_c({\vec {r^\prime}})$.
$${\vec j}_c({\vec {r^\prime}})=\int {d{\vec q}\over (2\pi)^2}
\exp{(i{\vec q}\cdot
{\vec {r^{\prime}}})}{2\pi\lambda_L^2\over i q^2c^2}
{\vec q}({\v\cdot{\vec q}}){\v\cdot{\hat z}\times{\vec q}\over
q^2\lambda_{TF}^2+1}\eqno{(A11)}$$
On performing the angular integration in Eq. (A11)
we arrive at the result of Eq. (19) where
$$I_3=\int dq {q^2 J_1(qr^{\prime})\over q^2\lambda_{TF}^2+1 }\eqno 
{(A12a)}$$
and
$$I_4=\int dq {q^2 J_3(qr^{\prime})\over q^2\lambda_{TF}^2+1} 
\eqno{(A12b)}.$$
$I_3$ is easily evaluated to yield
$$I_3={1\over \lambda_{TF}^3}K_1(r^{\prime}/\lambda_{TF})\eqno{(A13)}.$$
The Thomas-Fermi screening length $\lambda_{TF}\ll \xi$
where $\xi $ is the superconducting coherence length beyond which
our coarse-grained picture based on a phase-only functional
is valid. Thus $K_1(r^\prime/\lambda_{TF})\approx \exp{(-r^\prime/\lambda_{TF})}
\rightarrow 0$ in this ($r>\xi$)regime.

To evaluate $I_4$ we make use of the relation
$J_{m-1}(x) -J_{m+1}(x)=2J_{m}^{\prime}(x)$
which is obeyed by Bessel functions.
Using this relation, we find that $I_4$
can be rewritten as
$$I_4=\int dq {q^2 J_1(qr^{\prime})\over q^2\lambda_{TF}^2+1}
-2{\partial\over \partial r^{\prime}}\int dq {q [J_0(qr^\prime)
-2J_1^{\prime}(qr^\prime)]\over q^2\lambda_{TF}^2+1}\eqno{(A14)}.$$
This expression can be further simplified to yield
$$I_4={1\over \lambda_{TF}^3}K_1(r^\prime/\lambda_{TF})
-2{\partial\over\partial r^\prime }[{K_0(r^\prime /\lambda_{TF})\over
 \lambda_{TF}^2 }]+4{\partial^2\over \partial {r^\prime}^2}
 \int dq {J_1(qr^\prime)\over q^2\lambda_{TF}^2+1}\eqno{(A15)}. $$
 We ignore the first two terms on the right hand side
 of Eq. (A15) as they are exponentially small
 and in view of the small screening length, 
 approximate the last integral
 by
 $$\int dq {J_1(qr^\prime)\over (q^2\lambda_{TF}^2+1)}\approx
\int_0^{\xi^{-1}}dq J_1(qr^\prime)={1\over r^{\prime}}
 [1-J_0(r^\prime /\xi)]\eqno{(A16)}.$$
  Substituting these approximations in Eq. (A15)
  we finally arrive at the result of
  Eq. (20b).

\newpage
\noindent {\bf References}

\begin{enumerate}

\item G. Blatter, M.V. Feigel'man, V.B. Geshkenbein, A.I. Larkin and
V.M. Vinokur, {\em Rev. Mod. Phys.} {\bf 66}, 1125 (1994).

\item A. Oral et. al.,
{\em Phys. Rev. Lett.} {\bf 80}, 3610 (1998) and references
therein.

\item  A. F. Th. Hoekstra  et. al.,
{\em Phys. Rev. Lett.} {\bf 80}, 4293 (1998).

\item  T. Nagaoka   et. al.,
{\em Phys. Rev. Lett.} {\bf 80}, 3594 (1998);
S. Bhattacharya, M.J. Higgins and T.V. Ramakrishnan, {\em Phys. Rev.
Lett.} {\bf 73}, 1699 (1994).

\item S. Spielman et. al., {\em Phys. Rev. Lett.} {\bf 73}, 1537 (1994).

\item D. J. Thouless, P. Ao and Q. Niu
 {\em Phys. Rev. Lett.} {\bf 76}, 3758 (1996);
 G. E. Volovick JETP Lett. {\bf 62}, 66 (1995).
 
 \item D. M. Gaitonde and T. V. Ramakrishnan, Phys. Rev. {\bf B56},
 11,951 (1997) and references therein.
 
 \item J-M. Duan Phys. Rev. {\bf B48}, 333 (1993).
 
 \item A. van Otterlo, M. V. Feigelman,
 V. B. Geshkenbein and G. Blatter
  {\em Phys. Rev. Lett.} {\bf 75}, 3736 (1995).
 
 \item T. V. Ramakrishnan, {\em Physica Scripta} T {\bf 27}, 24 (1989);
  I. J. R. Aitchison et. al., {\em Phys. Rev.} B {\bf 51}, 6531 (1995).

\end{enumerate}

\end{document}